\documentclass[a4paper,11pt]{article}
\pdfoutput=1 

\usepackage{jcappub} 

\usepackage[T1]{fontenc} 

\usepackage{amssymb}
\usepackage{graphics}
\usepackage{latexsym}
\usepackage{graphicx}
\usepackage{bm}
\usepackage{amsmath}
\usepackage{float}
\usepackage{graphicx}
\usepackage{setspace}
\usepackage{amsfonts}
\usepackage{fancyhdr}
\usepackage{layout}
\usepackage{epstopdf}
\usepackage{graphicx}
\usepackage{minitoc}
\usepackage{slashed}
\usepackage{mathrsfs}
\usepackage{eufrak}
\usepackage{mathtools}
\usepackage[hang,flushmargin]{footmisc} 
\usepackage[sans,nouppercase]{frontespizio}
 
\usepackage{amsmath,mathtools}
\usepackage[ddmmyyyy]{datetime}
\usepackage{soul}
\usepackage[normalem]{ulem}
\usepackage{cancel}

\usepackage{hyperref}
\usepackage{color}
\usepackage{comment}
\usepackage{marvosym} 
\usepackage{amsmath,graphicx}
\usepackage[normalem]{ulem}
\usepackage{soul}
\usepackage{tcolorbox}
\newtcolorbox{alertblock}[1]{
colbacktitle=red!70!black, coltitle=white, fonttitle=\bfseries, title=#1,
colback=red!10!white, coltext=black,
colframe=red!70!black, boxrule=0.2pt
}
\newtcolorbox{block}[1]{
colbacktitle=blue!70!black, coltitle=white, fonttitle=\bfseries, title=#1,
colback=blue!10!white, coltext=black,
colframe=blue!70!black, boxrule=0.2pt
}
\usepackage{caption}
\usepackage{amssymb} 
\usepackage{tikz, tikz-cd} 
\usetikzlibrary{shapes.geometric, arrows}
\usepackage{tikz, tikz-cd}
\usetikzlibrary{shapes.geometric, arrows}
\definecolor{lightfuchsiapink}{rgb}{0.98, 0.52, 0.9} 
\tikzstyle{rec_red} = [rectangle, rounded corners, minimum width=3cm, minimum height=1cm,text centered, draw=black, fill=red!20]
\tikzstyle{rec_lightfuchsiapink} = [rectangle, rounded corners, minimum width=3cm, minimum height=1cm,text centered, draw=black, fill=lightfuchsiapink!20]
\tikzstyle{rec_blue} = [rectangle, rounded corners, minimum width=3cm, minimum height=1cm,text centered, draw=black, fill=blue!20]
\tikzstyle{rec_green} = [rectangle, rounded corners, minimum width=3cm, minimum height=1cm,text centered, draw=black, fill=green!20]
\tikzstyle{arrow} = [thick,*->>,>=stealth] 
\tikzset{
  shift left/.style ={commutative diagrams/shift left={#1}},
  shift right/.style={commutative diagrams/shift right={#1}}
}

\def\a{\alpha}
\def\r{\rho}
\def\s{\sigma}
\def\t{\tau}
\def\m{\mu}
\def\n{\nu}
\def\k{\kappa}
\def\th{\theta}
\def\g{\gamma}\def\G{\Gamma}
\def\L{\Lambda}\def\l{\lambda}
\def\D{\Delta}
\def\la{\langle}
\def\ra{\rangle}
\def\o{\omega}\def\O{\Omega}
\def\d{\delta}

\def\vp{\varphi}

\def\half{\textstyle{\frac{1}{2}}}

\def\bdoc{\begin{document}}
\def\edoc{\end{document}}

\def\beq{\begin{equation}}
\def\eeq{\end{equation}}
\def\bea{\begin{eqnarray}}
\def\eea{\end{eqnarray}}
\def\ben{\begin{enumerate}}
\def\een{\end{enumerate}}
\def\la{\langle}
\def\ra{\rangle}
\def\a{\alpha}
\def\b{\beta}
\def\g{\gamma}\def\G{\Gamma}
\def\d{\delta}\def\D{\Delta}
\def\e{\epsilon}
\def\z{\zeta}

\def\th{\theta}
\def\k{\kappa}
\def\l{\lambda}
\def\m{\mu}
\def\n{\nu}
\def\o{\omega}
\def\r{\rho}
\def\s{\sigma}
\def\t{\tau}
\def\L{{\cal L}}
\def\S{\Sigma }
\def\gsim{\; \raisebox{-.8ex}{$\stackrel{\textstyle >}{\sim}$}\;}
\def\lsim{\; \raisebox{-.8ex}{$\stackrel{\textstyle <}{\sim}$}\;}
\def\gtrsim{\gsim}
\def\lessim{\lsim}
\def\loc{{\rm local}}
\def\vm{v_{\rm max}}
\def\bh{\bar{h}}
\def\del{\partial}
\def\nab{\nabla}
\def\half{{\textstyle{\frac{1}{2}}}}
\def\fourth{{\textstyle{\frac{1}{4}}}}

\def\O{\Omega}

\def\br{{\bf r}}
\def\bnab{{\bf \nab}}

\def\tE{\tilde{E}}
\def\tL{\tilde{L}}
\def\Horava{Ho\v{r}ava }

\def\nn{\nonumber}

\title{\boldmath Cosmic structures and gravitational waves in ghost-free scalar-tensor theories of gravity}

\author[a,b,c]{Nicola Bartolo,}
\author[a,b]{Purnendu Karmakar,}
\author[a,b,c,d]{Sabino Matarrese}
\author[a]{and Mattia Scomparin}

\affiliation[a]{Dipartimento di Fisica e Astronomia "G. Galilei", Universit\`a degli Studi di Padova,\\via Marzolo 8, I-35131, Padova, Italy}
\affiliation[b]{INFN, Sezione di Padova,\\via F. Marzolo 8, 35131, Padova, Italy}
\affiliation[c]{INAF-Osservatorio Astronomico di Padova,\\vicolo dell'Osservatorio 5, I-35122, Padova, Italy}
\affiliation[d]{Gran Sasso Science Institute,\\via F. Crispi 7, I-67100, L'Aquila, Italy}

\emailAdd{nicola.bartolo@pd.infn.it}
\emailAdd{purnendu.karmakar@pd.infn.it}
\emailAdd{sabino.matarrese@pd.infn.it}
\emailAdd{mattia.scompa@gmail.com}
\abstract{
We study cosmic structures in the quadratic Degenerate Higher Order Scalar Tensor (qDHOST) model, which has been proposed as the most general scalar-tensor theory 
(up to quadratic dependence on the covariant derivatives of the scalar field), which is not plagued by the presence of ghost instabilities. 
We then study a static, spherically symmetric object embedded in de Sitter space-time for the qDHOST model. This model exhibits breaking of the Vainshtein mechanism inside the cosmic structure and Schwarzschild-de Sitter space-time outside, where General Relativity (GR) can be recovered within the Vainshtein radius. We then look for the conditions on the parameters on the considered qDHOST scenario which ensure the validity of the Vainshtein screening mechanism inside the object and the fulfilment of the recent GW170817/GRB170817A constraint on the speed of propagation of gravitational waves. We find that these two constraints rule out the same set of parameters, corresponding to the Lagrangians that are quadratic in second-order derivatives of the scalar field, for the shift symmetric qDHOST.
}

\newcommand\ddfrac[2]{\frac{\displaystyle #1}{\displaystyle #2}}

\begin{document}
\maketitle
\flushbottom

\allowdisplaybreaks

\section{Introduction} \label{sec:intro}
Modelling the recent cosmic acceleration phase of the universe expansion \cite{Perlmutter:1998np, Riess:1998cb, Hinshaw:2012aka}
has become one of the greatest challenges of modern theoretical cosmology. A very useful general framework in this direction is provided by 
scalar-tensor gravity theories, where GR is extended by introducing one or more scalar degrees of freedom. 
A necessary requirement for any such extension is that it should not introduce higher-order time derivatives in the equation of motion, known as Ostrogradsky theorem \cite{Woodard:2006nt,Burgess:2014lwa,Woodard:2015zca}. 
The Horndeski or ``generalized Galileon'' theory \cite{Horndeski:1974wa} was originally proposed as the most general Ostrogradsky ghost-free scalar-tensor theory. The authors of Ref. \cite{Gleyzes:2014dya,Gleyzes:2014qga} claimed that their extension of the Horndeski theory to the so-called Gleyzes-Langlois-Piazza-Vernizzi (GLPV) theory, leads to a new class of models without the Ostrogradsky ghost instability. 
A later proposal showed that higher-order derivatives in the Lagrangian may not necessarily introduce Ostrogradsky ghosts, provided certain degeneracy conditions are met \cite{Langlois:2015cwa}. Indeed, ``degenerate'' Lagrangians with non-invertible kinetic matrix will ensure that the number of degrees of freedom is preserved, thus
making the theory free from the Ostrogradsky ghost; this new class was named ``degenerate higher order scalar-tensor" (DHOST) theory \cite{Crisostomi:2016czh,Achour:2016rkg,BenAchour:2016fzp}. A particular extension of Horndeski was proposed earlier in Ref. \cite{Zumalacarregui:2013pma}, which appeared as a result of a disformal transformation on the Einstein-Hilbert action, later found a specific subclass of HOST theory. DHOST theory is categorised into several classes \cite{BenAchour:2016fzp}. Class I DHOST theories are the only one which are healthy from the gradient instability, i.e., the square of the speed of the tensor modes (gravitational-wave speed) and that of the scalar mode (sound speed) do not have opposite 
sign, $c_s^2 \propto - {c_T}^2$ \cite{Langlois:2017mxy}. 

Gravity is well tested and established on small scales (e.g. laboratory, solar system, ...). Therefore, there must be a screening mechanism able to suppress the fifth-force mediated by the new scalar degree of freedom, without destroying the modifications on large scales, while recovering GR on a small scale. 
In general, the so-called Vainshtein screening is widely used for higher-order scalar-tensor theories \cite{Vainshtein:1972sx}. 
In the Vainshtein screening, the non-linear self-interactions of the scalar field suppress the propagation of the fifth-force near the matter source \cite{Babichev:2013usa,Brax:2004qh}. The Vainshtein mechanism in the Horndeski framework has been studied intensively in Refs. \cite{Kimura:2011dc,Narikawa:2013pjr,Koyama:2013paa,Kase:2013uja}. Although standard gravity is recovered outside a non-relativistic static and spherically symmetric cosmic structure in the small scale limit, the Vainshtein mechanism breaks down inside the matter source for the specific case of GLPV-beyond Horndeski models \cite{Kobayashi:2014ida,Koyama:2015oma, Saito:2015fza,Sakstein:2015zoa, Sakstein:2015aac,Jain:2015edg,Sakstein:2016ggl,Babichev:2016jom}. The consequences of these predictions for
astrophysical and cosmological observations have been discussed in \cite{Sakstein:2016oel,Sakstein:2017xjx}. 
A similar breakdown of the Vainshtein mechanism in some classes of the qDHOST model has been studied in some recent articles 
\cite{Crisostomi:2017lbg, Langlois:2017dyl}.

The recent multi-messenger gravitational wave (GW) event, GW170817 \cite{TheLIGOScientific:2017qsa} by the LIGO/VIRGO collaboration and the associate gamma-ray burst (GRB) event, GRB170817A \cite{Monitor:2017mdv} put a very tight constraint on the speed of gravitational wave propagation with respect to the speed of light, $|c_T^2/c^2 - 1|\le 5 \times 10^{-16}$. This constraint is so narrow that we may generically consider $c_T^2 = c^2$. Many of the aforementioned Horndeski and beyond Horndeski models predict significant deviations in the speed of GW from the speed of light \cite{deRham:2016wji}. Constraining the scalar-tensor theories in the light of the anomalous speed of GW propagation have been first studied in Ref. \cite{Lombriser:2015sxa}, and followed by many recent articles \cite{Bettoni:2016mij, Ezquiaga:2017ekz, Sakstein:2017xjx,Creminelli:2017sry,Baker:2017hug, Jain:2015edg, Arai:2017hxj}.

Here we focus on the specific subclass of class I of qDHOST models for simplicity, where the degeneracy conditions are on the scalar sector alone, which does not suffer from the gradient and ghost instability \cite{Langlois:2017mxy}. We shall study the `Vainshtein mechanism' in Ia* class of models in the presence of spherically symmetric cosmic structures. In section \ref{sec:action}, we briefly discuss the shift-symmetric qDHOST theory of gravity. The covariant field equations of qHOST theories, therefore including all qDHOST classes, are derived in section \ref{sec:eom:host}. In the following section, section \ref{sec:back:dS}, we discuss the field equations in a de Sitter background. In section \ref{sec:pert:qdhost} we derive the perturbed equations around the de Sitter background for a static spherically symmetric matter source, taking the sub-Horizon, non-relativistic weak-field limit. 
We thus obtain the explicit form of the modified Newton's constant and the two gravitational potentials. In section \ref{sec:gw:dhost} we compute 
the speed of GW propagation for that model and show that the speed of GWs and the screening mechanism are governed by the same functional parameters for this class of models. Finally, section \ref{sec:concl} presents our main conclusions.
We use the metric signature $(-,+,+,+)$ and we set the speed of light and reduced Planck mass to unity. Greek indices run from 0 to 3. 


\section{The qDHOST theory}
\label{sec:action}
Let us consider the action for the shift-symmetric quadratic higher-order scalar tensor theory, which is expressed as follows \cite{Achour:2016rkg}, 
\begin{eqnarray}\label{S}
 S&=& \int d^4x\,  \sqrt{-g}\,  \mathcal{L}\,, \label{ac:total}
\end{eqnarray}
where the total Lagrangian, $\mathcal{L}$ is defined as the sum of the following four parts,
\begin{equation}\label{l:total}
\mathcal{L} = \mathcal{L}_g + \mathcal{L}_\vp + \mathcal{L}_{oth} + \mathcal{L}_m \,,
\end{equation}
with
\begin{eqnarray} \label{l:exp}
\mathcal{L}_g &\equiv& f R\,,\\
\mathcal{L}_{\vp}&\equiv&\sum_{I=1}^5\zeta_{I} (X) \mathcal{L}_{I}\,,
\\
\mathcal{L}_{oth}&\equiv&\big(A\,X- B\Lambda\big) \,,
\end{eqnarray}
where $g$ is the determinant of the metric $g_{\m\n}$. The Lagrangian $\mathcal{L}_g$ is built in terms of the metric tensor and Ricci scalar, $\mathcal{L}_\vp$ is the second-order contraction of the derivatives of a scalar field, $\vp$; $\mathcal{L}_{oth}$ are the terms which do not affect the degeneracy condition, and $\mathcal{L}_m$ is the matter Lagrangian coupled only with the metric $g_{\m\n}$. For simplicity, we have considered the free coefficients $f,A, B$ to be constant, $\Lambda$ is a (positive) cosmological constant \footnote{The dimensions of the introduced parameters are as follows: $[A]=[B]=[M]^0$, $[f]=[M]^2$, and $[\zeta_1]=[\zeta_2]=[M]^{-2}$ and $[\zeta_3]=[\zeta_4]=[M]^{-6}$, $[\zeta_5]=[M]^{-10}$, $[\Lambda]=[M]^4$ and $[H]=[M]$.
}. The above action is shift-symmetric with respect to the scalar field, 
$\vp \rightarrow \vp + const$. Therefore all $\vp$ contributions and dependences appear only as contractions of first and second-order covariant derivatives.

The five arbitrary functions, $\zeta_I=\zeta_{I}(X)$ depend only on the kinetic term, $X\equiv \nabla_\mu\vp\nabla^\mu\vp$. In our notation the $\nabla_\mu$ symbol indicates the covariant derivative.

The five possible quadratic dependences on the covariant derivatives of the scalar field, $\vp$, are, $\mathcal{L}_I$ \cite{Achour:2016rkg},
\begin{align}\label{l:p:exp}
\mathcal{L}_1 &\equiv\nabla^{\mu}\nabla^\nu\vp\nabla_{\mu}\nabla_\nu\vp\,, \nonumber\\
\mathcal{L}_2 &\equiv  (\Box \vp)^2\,,\nonumber\\
\mathcal{L}_3 &\equiv  (\Box \vp )\nabla^{\mu}\vp \nabla^{\nu}\vp \nabla_{\mu}\nabla_{ \nu}\vp\,, \nonumber \\
\mathcal{L}_4&\equiv \nabla^{\mu}\vp \nabla^{\nu}\vp \nabla_{\mu}\nabla^{\rho}\vp \nabla_{\nu}\nabla_{ \rho}\vp\,,\nonumber\\
\mathcal{L}_5 &\equiv (\nabla^{\mu} \vp \nabla_{\mu}\nabla_{ \nu}\vp \nabla^{\nu}\vp)^2\,.
\end{align}

The vanishing of the determinant of the kinetic matrix leads to the three different degeneracy conditions associated with second-class constraints, which lead to the seven classes of the ghost-free degenerate theories within the qHOST \cite{Achour:2016rkg}. We will study the degenerate subclass characterised by the degeneracy of the scalar sector alone, and we call this class Ia*. 
The condition on the coefficients for such a ghost-free subclass is given as follows,
\begin{equation}\label{d:dhost}
\zeta_2(X)=-\zeta_1(X)\qquad\quad\zeta_3(X)=-\zeta_4(X)=2X^{-1}\zeta_1(X)\qquad\quad\zeta_5(X)=0\,.
\end{equation}
$\zeta_5(X)=0$ is in general non-zero in the general zero for the class I qDHOST. It turns out to be zero for this restricted class of model.

\section{Covariant field equations of qHOST theories} \label{sec:eom:host}

In this section, we will obtain the explicit expression of the covariant field equations for the qHOST theory. 
The scalar field equation of the qHOST action, Eq. (\ref{ac:total}), is $\delta S / \delta\varphi=0$. Considering that our action is shift-symmetric, the scalar field equation leads to the 
current-conservation law
\begin{equation} \label{eom:phi}
\nabla_\mu\mathcal{J}^\mu=0\,,
\end{equation}
where the $\mathcal{J}$ is the current for qHOST theories, whose expression is 
\begin{eqnarray}\label{eom:phi:exp}
\mathcal{J}^{\mu} &=&\Big\{-2A \nabla^{\mu} \varphi\Big\}_{(oth)}\nonumber\\
&+&\Big\{\zeta_{1X}\big(4\nabla^{\nu} \varphi \nabla^{\mu} \nabla^{\rho} \varphi \nabla_{\nu} \nabla_{\rho} \varphi-2\nabla^{\mu} \varphi \nabla^{\nu} \nabla^{\rho} \varphi \nabla_{\nu} \nabla_{\rho} \varphi\big)+\zeta_{1} \big(2\nabla^{\nu} \nabla_{\nu} \nabla^{\mu} \varphi\big)\Big\}_{(1)}\nonumber\\
&+&\Big\{\zeta_{2X} \big(4\nabla^{\nu} \varphi \nabla^{\mu} \nabla_{\nu} \varphi \nabla^{\rho} \nabla_{\rho} \varphi-2\nabla^{\mu} \varphi \nabla^{\nu} \nabla_{\nu} \varphi \nabla^{\rho} \nabla_{\rho} \varphi\big)+\zeta_{2}\big(2 \nabla^{\mu} \nabla^{\nu}\,{\nabla}_{\nu} \varphi\big)\Big\}_{(2)}\nonumber\\
&+&\Big\{\zeta_{3X} \big(2\nabla^{\nu} \varphi \nabla^{\rho} \varphi \nabla^{\sigma} \varphi \nabla^{\mu} \nabla_{\nu} \varphi \nabla_{\rho} \nabla_{\sigma} \varphi\big)+\zeta_{3} \big(2\nabla^{\nu} \varphi \nabla^{\mu} \nabla^{\rho} \varphi \nabla_{\nu} \nabla_{\rho} \varphi+\nabla^{\nu} \varphi \nabla^{\rho} \varphi \nabla^{\mu} \nabla_{\nu} \nabla_{\rho} \varphi\nonumber\\
&&+\,\nabla^{\mu} \varphi \nabla^{\nu} \nabla_{\nu} \varphi \nabla^{\rho} \nabla_{\rho} \varphi-\nabla^{\nu} \varphi \nabla^{\mu} \nabla_{\nu} \varphi \nabla^{\rho} \nabla_{\rho} \varphi+\nabla^{\mu} \varphi \nabla^{\nu} \varphi \nabla_{\nu} \nabla^{\rho} \nabla_{\rho} \varphi\big)\Big\}_{(3)}\nonumber\\
&+&\Big\{\zeta_{4X}\big(2 \nabla^{\nu} \varphi \nabla^{\rho} \varphi \nabla^{\sigma} \varphi \nabla^{\mu} \nabla_{\nu} \varphi \nabla_{\rho} \nabla_{\sigma} \varphi\big)+\zeta_{4} \big(\nabla^{\nu} \varphi \nabla^{\mu} \nabla_{\nu} \varphi \nabla^{\rho} \nabla_{\rho} \varphi+\nabla^{\nu} \varphi \nabla^{\rho} \varphi \nabla_{\nu} \nabla^{\mu} \nabla_{\rho} \varphi\nonumber\\
&&+\,\nabla^{\mu} \varphi \nabla^{\nu} \nabla^{\rho} \varphi \nabla_{\nu} \nabla_{\rho} \varphi+\nabla^{\mu} \varphi \nabla^{\nu} \varphi \nabla^{\rho} \nabla_{\rho} \nabla_{\nu} \varphi\big)\Big\}_{(4)}\nonumber\\
&+&\Big\{\zeta_{5X}\big(2 \nabla^{\mu} \varphi \nabla^{\alpha} \varphi \nabla^{\nu} \varphi \nabla^{\rho} \varphi \nabla^{\sigma} \varphi \nabla_{\alpha} \nabla_{\nu} \varphi \nabla_{\rho} \nabla_{\sigma} \varphi\big)+\zeta_{5} \big(2\nabla^{\mu} \varphi \nabla^{\nu} \varphi \nabla^{\rho} \varphi \nabla_{\nu} \nabla_{\rho} \varphi \nabla^{\sigma} \nabla_{\sigma} \varphi\nonumber\\[1.0ex]
&&-\,2\nabla^{\nu} \varphi \nabla^{\rho} \varphi \nabla^{\sigma} \varphi \nabla^{\mu} \nabla_{\nu} \varphi \nabla_{\rho} \nabla_{\sigma} \varphi+4\nabla^{\mu} \varphi \nabla^{\nu} \varphi \nabla^{\rho} \varphi \nabla_{\nu} \nabla^{\sigma} \varphi \nabla_{\rho} \nabla_{\sigma} \varphi\nonumber\\[1.0ex]
&&+\,2\nabla^{\mu} \varphi \nabla^{\nu} \varphi \nabla^{\rho} \varphi \nabla^{\sigma} \varphi \nabla_{\nu} \nabla_{\rho} \nabla_{\sigma} \varphi\big)\Big\}_{(5)}\,,
\end{eqnarray}
The subscripts $(g)$, number $(I)$, $(oth)$, and $(m)$ in parentheses indicate the correspondence with the term of the Lagrangian, $\mathcal{L}_{g}$, $\mathcal{L}_I$, $\mathcal{L}_{others}$ and $\mathcal{L}_m$.

The equation of motion of qHOST with respect to the metric field, $g_{\mu\nu}$, is $\delta S / \delta g^{\mu\nu}=0$, 
\begin{equation}\label{eom:metric}
\mathcal{H}_{\mu\nu}=0\,, 
\end{equation}
with 
\begin{eqnarray}\label{eom:metric:exp}
\mathcal{H}_{\mu\nu} &\equiv& \Big\{2f G_{\mu\nu}\Big\}_{(g)}-\Big\{T_{\mu\nu}\Big\}_{(m)}+\Big\{A \big(2\nabla_{\mu} \varphi \nabla_{\nu} \varphi-g_{\mu\nu} X\big)+B  \Lambda g_{\mu\nu}\Big\}_{(oth)}\nonumber\\
&+&\Big\{\zeta_{1X} \big(2\nabla_{\mu} \varphi \nabla_{\nu} \varphi \nabla^{\rho} \nabla^{\sigma} \varphi \nabla_{\rho} \nabla_{\sigma} \varphi+4\nabla^{\rho} \varphi \nabla^{\sigma} \varphi \nabla_{\mu} \nabla_{\nu} \varphi \nabla_{\rho} \nabla_{\sigma} \varphi-8\nabla_{\mu} \varphi \nabla^{\rho} \varphi \nabla_{\nu} \nabla^{\sigma} \varphi \nabla_{\rho} \nabla_{\sigma} \varphi \nonumber\\
&&+\,\zeta_{1} \big(2\nabla_{\mu} \nabla_{\nu} \varphi \nabla^{\rho} \nabla_{\rho} \varphi+2\nabla^{\rho} \varphi \nabla_{\rho} \nabla_{\mu} \nabla_{\nu} \varphi
-4\nabla_{\mu} \varphi \nabla^{\rho} \nabla_{\rho} \nabla_{\nu} \varphi
-g_{\mu\nu} \nabla^{\rho} \nabla^{\sigma} \varphi \nabla_{\rho} \nabla_{\sigma} \varphi\big)\Big\}_{(1)}\nonumber\\
&+&\Big\{\zeta_{2X} \big(2\nabla_{\mu} \varphi \nabla_{\nu} \varphi \nabla^{\rho} \nabla_{\rho} \varphi \nabla^{\sigma} \nabla_{\sigma} \varphi+4g_{\mu\nu} \nabla^{\alpha} \varphi \nabla^{\rho} \varphi \nabla_{\alpha} \nabla_{\rho} \varphi \nabla^{\sigma} \nabla_{\sigma} \varphi\nonumber\\
&&-\,8\nabla_{\mu} \varphi \nabla^{\rho} \varphi \nabla_{\nu} \nabla_{\rho} \varphi \nabla^{\sigma} \nabla_{\sigma} \varphi\big)+\zeta_{2} \big(g_{\mu\nu} \nabla^{\rho} \nabla_{\rho} \varphi \nabla^{\sigma} \nabla_{\sigma} \varphi+2g_{\mu\nu} \nabla^{\rho} \varphi \nabla_{\rho} \nabla^{\sigma} \nabla_{\sigma} \varphi\nonumber\\
&&-\,4\nabla_{\mu} \varphi \nabla^\nu\nabla^\rho\nabla_\rho\vp \big)\Big\}_{(2)}\nonumber\\
&+&\Big\{\zeta_{3X} \big(2g_{\mu\nu} \nabla^{\alpha} \varphi \nabla^{\beta} \varphi \nabla^{\rho} \varphi \nabla^{\sigma} \varphi \nabla_{\alpha} \nabla_{\beta} \varphi \nabla_{\rho} \nabla_{\sigma} \varphi-4\nabla_{\mu} \varphi \nabla^{\alpha} \varphi \nabla^{\rho} \varphi \nabla^{\sigma} \varphi \nabla_{\nu} \nabla_{\alpha} \varphi \nabla_{\rho} \nabla_{\sigma} \varphi\big)\nonumber\\
&&+\,\zeta_{3} \big(2g_{\mu\nu} \nabla^{\alpha} \varphi \nabla^{\rho} \varphi \nabla_{\alpha} \nabla^{\sigma} \varphi \nabla_{\rho} \nabla_{\sigma} \varphi+g_{\mu\nu} \nabla^{\alpha} \varphi \nabla^{\rho} \varphi \nabla^{\sigma} \varphi \nabla_{\alpha} \nabla_{\rho} \nabla_{\sigma} \varphi\nonumber\\
&&+\,2\nabla_{\nu} \varphi \nabla^{\rho} \varphi \nabla_{\mu} \nabla_{\rho} \varphi \nabla^{\sigma} \nabla_{\sigma} \varphi-4\nabla_{\mu} \varphi \nabla^{\rho} \varphi \nabla_{\nu} \nabla^{\sigma} \varphi \nabla_{\rho} \nabla_{\sigma} \varphi-2\nabla_{\mu} \varphi \nabla^{\rho} \varphi \nabla^{\sigma} \varphi \nabla_{\nu} \nabla_{\rho} \nabla_{\sigma} \varphi\nonumber\\
&&-\,\nabla_{\nu} \varphi \nabla_{\mu} \varphi \nabla^{\rho} \nabla_{\rho} \varphi \nabla^{\sigma} \nabla_{\sigma} \varphi-\nabla_{\nu} \varphi \nabla_{\mu} \varphi \nabla^{\rho} \varphi \nabla_{\rho} \nabla^{\sigma} \nabla_{\sigma} \varphi\big)\Big\}_{(3)}\nonumber\\
&+&\Big\{\zeta_{4X}\big(-2 \nabla_{\nu} \varphi \nabla_{\mu} \varphi \nabla^{\alpha} \varphi \nabla^{\rho} \varphi \nabla_{\alpha} \nabla^{\sigma} \varphi \nabla_{\rho} \nabla_{\sigma} \varphi\big)+\zeta_{4} \big(2\nabla^{\rho} \varphi \nabla^{\sigma} \varphi \nabla_{\mu} \nabla_{\rho} \varphi \nabla_{\nu} \nabla_{\sigma} \varphi\nonumber\\
&&-\,2\nabla_{\mu} \varphi \nabla_{\nu} \varphi \nabla^{\rho} \nabla^{\sigma} \varphi \nabla_{\rho} \nabla_{\sigma} \varphi-2\nabla_{\nu} \varphi \nabla_{\mu} \varphi \nabla^{\rho} \varphi \nabla^{\sigma} \nabla_{\sigma} \nabla_{\rho} \varphi\nonumber\\
&&-\,g_{\mu\nu} \nabla^{\alpha} \varphi \nabla^{\rho} \varphi \nabla_{\alpha} \nabla^{\sigma} \varphi \nabla_{\rho} \nabla_{\sigma} \varphi\big)\Big\}_{(4)}\nonumber\\
&+&\Big\{\zeta_{5X} \big(-2\nabla_{\mu} \varphi \nabla_{\nu} \varphi \nabla^{\alpha} \varphi \nabla^{\beta} \varphi \nabla^{\rho} \varphi \nabla^{\sigma} \varphi \nabla_{\alpha} \nabla_{\beta} \varphi \nabla_{\rho} \nabla_{\sigma} \varphi\nonumber\\
&&+\,\zeta_{5} \big(4\nabla_{\nu} \varphi \nabla^{\alpha} \varphi \nabla^{\rho} \varphi \nabla^{\sigma} \varphi \nabla_{\mu} \nabla_{\alpha} \varphi \nabla_{\rho} \nabla_{\sigma} \varphi-2\nabla_{\nu} \varphi \nabla_{\mu} \varphi \nabla^{\alpha} \varphi \nabla^{\rho} \varphi \nabla_{\alpha} \nabla_{\rho} \varphi \nabla^{\sigma} \nabla_{\sigma} \varphi\nonumber\\
&&-\,4\nabla_{\nu} \varphi \nabla_{\mu} \varphi \nabla^{\alpha} \varphi \nabla^{\rho} \varphi \nabla_{\alpha} \nabla^{\sigma} \varphi \nabla_{\rho} \nabla_{\sigma} \varphi-2\nabla_{\nu} \varphi \nabla_{\mu} \varphi \nabla^{\alpha} \varphi \nabla^{\rho} \varphi \nabla^{\sigma} \varphi \nabla_{\alpha} \nabla_{\rho} \nabla_{\sigma} \varphi\nonumber\\
&&-\,g_{\mu\nu} \nabla^{\alpha} \varphi \nabla^{\beta} \varphi \nabla^{\rho} \varphi \nabla^{\sigma} \varphi \nabla_{\alpha} \nabla_{\beta} \varphi \nabla_{\rho} \nabla_{\sigma} \varphi\big)\Big\}_{(5)}\,,
\end{eqnarray}
where $G_{\mu\nu}$ is the standard Einstein tensor and the energy-momentum tensor $T_{\m\n}$ is defined as,
\begin{eqnarray}
T^{\mu\nu}&\equiv&\frac{2}{\sqrt{-g}}\frac{\delta (\sqrt{-g}\mathcal{L}_m)}{\delta g_{\mu\nu}}\,.
\end{eqnarray}

These field equations, (\ref{eom:phi}) and (\ref{eom:metric}) in general contain higher-order derivatives of the fields \cite{BenAchour:2016fzp}. 
The proper degeneracy condition on $f$ and $\zeta_I$ leads to the field equations corresponding to all the different classes of the qDHOST theories. As explained in the previous section, we have restricted our analysis to the Ia* qDHOST subclass, defined by Eq. (\ref{d:dhost}).

\subsection{Covariant field equations of qDHOST theories in degenerate subclass Ia*} \label{sec:eom:qdhost}

One can derive the field equations of the Ia* qDHOST subclass from Eqs. \eqref{eom:phi:exp} and \eqref{eom:metric:exp}, by applying the degeneracy conditions given in Eq. \eqref{d:dhost}.
\begin{equation}\label{Ia*qDHOST}
\nabla J^\mu=0\,,\qquad\qquad H_{\mu\nu}=0\,.
\end{equation}
To formally distinguish the qDHOST EOMs of class Ia* from qHOST EOMs, we change the notation $\mathcal{J}^\mu\rightarrow J^\mu$ and $\mathcal{H}_{\mu\nu}\rightarrow H_{\mu\nu}$, where
\begin{eqnarray}
J^{\mu} &=& -2A \nabla^{\mu} \varphi+\zeta_{1X} \big(4\nabla^{\nu} \varphi \nabla^{\mu} \nabla^{\rho} \varphi \nabla_{\nu} \nabla_{\rho} \varphi-2\nabla^{\mu} \varphi \nabla^{\nu} \nabla^{\rho} \varphi \nabla_{\nu} \nabla_{\rho} \varphi-4\nabla^{\nu} \varphi \nabla^{\mu} \nabla_{\nu} \varphi \nabla^{\rho} \nabla_{\rho} \varphi\nonumber\\[1.0ex]
&+&2\nabla^{\mu} \varphi \nabla^{\nu} \nabla_{\nu} \varphi \nabla^{\rho} \nabla_{\rho} \varphi\big)+\zeta_{1} \big(2\nabla^{\nu} \nabla_{\nu} \nabla^{\mu} \varphi-2\nabla^{\mu} \nabla^{\nu}{\nabla}_{\nu} \varphi\big)+\zeta_{1}X^{-1} \big(4\nabla^{\nu} \varphi \nabla^{\mu} \nabla^{\rho} \varphi \nabla_{\nu} \nabla_{\rho} \varphi\nonumber\\[1.0ex]
&+&2\nabla^{\nu} \varphi \nabla^{\rho} \varphi \nabla^{\mu} \nabla_{\nu} \nabla_{\rho} \varphi+2\nabla^{\mu} \varphi \nabla^{\nu} \nabla_{\nu} \varphi \nabla^{\rho} \nabla_{\rho} \varphi-4\nabla^{\nu} \varphi \nabla^{\mu} \nabla_{\nu} \varphi \nabla^{\rho} \nabla_{\rho} \varphi\nonumber\\[1.0ex]
&+&2\nabla^{\mu} \varphi \nabla^{\nu} \varphi \nabla_{\nu} \nabla^{\rho} \nabla_{\rho} \varphi-2\nabla^{\nu} \varphi \nabla^{\rho} \varphi \nabla_{\nu} \nabla^{\mu} \nabla_{\rho} \varphi-2\nabla^{\mu} \varphi \nabla^{\nu} \nabla^{\rho} \varphi \nabla_{\nu} \nabla_{\rho} \varphi\nonumber\\[1.0ex]
&-&2\nabla^{\mu} \varphi \nabla^{\nu} \varphi \nabla^{\rho} \nabla_{\rho} \nabla_{\nu} \varphi\big)\,,
\end{eqnarray}
and where
\begin{eqnarray}
H_{\mu\nu} &=& 2f G_{\mu\nu}+A \big(2\nabla_{\mu} \varphi \nabla_{\nu} \varphi-g_{\mu\nu} X\big)+ B\Lambda g_{\mu\nu} -T_{\mu\nu}+\zeta_{1X} \big(2\nabla_{\mu} \varphi \nabla_{\nu} \varphi \nabla^{\rho} \nabla^{\sigma} \varphi \nabla_{\rho} \nabla_{\sigma} \varphi\nonumber\\[1.0ex]
&+&4\nabla^{\rho} \varphi \nabla^{\sigma} \varphi \nabla_{\mu} \nabla_{\nu} \varphi \nabla_{\rho} \nabla_{\sigma} \varphi-8\nabla_{\mu} \varphi \nabla^{\rho} \varphi \nabla_{\nu} \nabla^{\sigma} \varphi \nabla_{\rho} \nabla_{\sigma} \varphi-2\nabla_{\mu} \varphi \nabla_{\nu} \varphi \nabla^{\rho} \nabla_{\rho} \varphi \nabla^{\sigma} \nabla_{\sigma} \varphi\nonumber\\
&-&4g_{\mu\nu} \nabla^{\alpha} \varphi \nabla^{\rho} \varphi \nabla_{\alpha} \nabla_{\rho} \varphi \nabla^{\sigma} \nabla_{\sigma} \varphi+8\nabla_{\mu} \varphi \nabla^{\rho} \varphi \nabla_{\nu} \nabla_{\rho} \varphi \nabla^{\sigma} \nabla_{\sigma} \varphi+\zeta_{1} \big(2\nabla^{\rho} \varphi \nabla_{\rho} \nabla_{\mu} \nabla_{\nu} \varphi\nonumber\\
&-&4\nabla_{\mu} \varphi \nabla^{\rho} \nabla_{\rho} \nabla_{\nu} \varphi-g_{\mu\nu} \nabla^{\rho} \nabla^{\sigma} \varphi \nabla_{\rho} \nabla_{\sigma} \varphi-\,g_{\mu\nu} \nabla^{\rho} \nabla_{\rho} \varphi \nabla^{\sigma} \nabla_{\sigma} \varphi-2g_{\mu\nu} \nabla^{\rho} \varphi \nabla_{\rho} \nabla^{\sigma} \nabla_{\sigma} \varphi\nonumber\\
&+&2\nabla_{\mu} \varphi \nabla_{\nu}\nabla^{\rho}\nabla_{\rho}+2\nabla_{\nu} \nabla_{\mu} \varphi \nabla^{\rho} \nabla_{\rho} \varphi+2\nabla_{\nu} \varphi \nabla_{\mu} \nabla^{\rho} \nabla_{\rho} \varphi\big)\nonumber\\
&+&\zeta_{1X}{X}^{-1} \big(4g_{\mu\nu} \nabla^{\alpha} \varphi \nabla^{\beta} \varphi \nabla^{\rho} \varphi \nabla^{\sigma} \varphi \nabla_{\alpha} \nabla_{\beta} \varphi \nabla_{\rho} \nabla_{\sigma} \varphi-8\nabla_{\mu} \varphi \nabla^{\alpha} \varphi \nabla^{\rho} \varphi \nabla^{\sigma} \varphi \nabla_{\nu} \nabla_{\alpha} \varphi \nabla_{\rho} \nabla_{\sigma} \varphi\nonumber\\[1.0ex]
&+&4\nabla_{\nu} \varphi \nabla_{\mu} \varphi \nabla^{\alpha} \varphi \nabla^{\rho} \varphi \nabla_{\alpha} \nabla^{\sigma} \varphi \nabla_{\rho} \nabla_{\sigma} \varphi\big)+\zeta_{1}{X}^{-2} \big(-4g_{\mu\nu} \nabla^{\alpha} \varphi \nabla^{\beta} \varphi \nabla^{\rho} \varphi \nabla^{\sigma} \varphi \nabla_{\alpha} \nabla_{\beta} \varphi \nabla_{\rho} \nabla_{\sigma} \varphi\nonumber\\[1.0ex]
&+&8\nabla_{\mu} \varphi \nabla^{\alpha} \varphi \nabla^{\rho} \varphi \nabla^{\sigma} \varphi \nabla_{\nu} \nabla_{\alpha} \varphi \nabla_{\rho} \nabla_{\sigma} \varphi-4\nabla_{\nu} \varphi \nabla_{\mu} \varphi \nabla^{\alpha} \varphi \nabla^{\rho} \varphi \nabla_{\alpha} \nabla^{\sigma} \varphi \nabla_{\rho} \nabla_{\sigma} \varphi\big)\nonumber\\
&+&\zeta_{1}{X}^{-1} \big(6g_{\mu\nu} \nabla^{\alpha} \varphi \nabla^{\rho} \varphi \nabla_{\alpha} \nabla^{\sigma} \varphi \nabla_{\rho} \nabla_{\sigma} \varphi+2g_{\mu\nu} \nabla^{\alpha} \varphi \nabla^{\rho} \varphi \nabla^{\sigma} \varphi \nabla_{\alpha} \nabla_{\rho} \nabla_{\sigma} \varphi\nonumber\\
&-&2\nabla_{\mu} \varphi \nabla^{\rho} \varphi \nabla^{\sigma} \varphi \nabla_{\nu} \nabla_{\rho} \nabla_{\sigma} \varphi+6\nabla_{\nu} \varphi \nabla^{\rho} \varphi \nabla_{\mu} \nabla_{\rho} \varphi \nabla^{\sigma} \nabla_{\sigma} \varphi-2\nabla_{\nu} \varphi \nabla^{\rho} \varphi \nabla^{\sigma} \varphi \nabla_{\mu} \nabla_{\rho} \nabla_{\sigma} \varphi\nonumber\\
&-&2\nabla_{\mu} \varphi \nabla^{\rho} \varphi \nabla_{\nu} \nabla_{\rho} \varphi \nabla^{\sigma} \nabla_{\sigma} \varphi-2\nabla_{\nu} \varphi \nabla_{\mu} \varphi \nabla^{\rho} \nabla_{\rho} \varphi \nabla^{\sigma} \nabla_{\sigma} \varphi\nonumber\\
&-&2\nabla_{\nu} \varphi \nabla_{\mu} \varphi \nabla^{\rho} \varphi \nabla_{\rho} \nabla^{\sigma} \nabla_{\sigma} \varphi-6\nabla^{\rho} \varphi \nabla^{\sigma} \varphi \nabla_{\mu} \nabla_{\rho} \varphi \nabla_{\nu} \nabla_{\sigma} \varphi-8\nabla_{\mu} \varphi \nabla^{\rho} \varphi \nabla_{\nu} \nabla^{\sigma} \varphi \nabla_{\rho} \nabla_{\sigma} \varphi\nonumber\\
&+&4\nabla_{\mu} \varphi \nabla_{\nu} \varphi \nabla^{\rho} \nabla^{\sigma} \varphi \nabla_{\rho} \nabla_{\sigma} \varphi+4\nabla_{\mu} \varphi \nabla_{\nu} \varphi \nabla^{\rho} \varphi \nabla^{\sigma} \nabla_{\sigma} \nabla_{\rho} \varphi+2\nabla^{\rho} \varphi \nabla^{\sigma} \varphi \nabla_{\nu} \nabla_{\rho} \varphi \nabla_{\mu} \nabla_{\sigma} \varphi\big)\,.\nn\\
\end{eqnarray}

\section{qDHOST in de Sitter background}
\label{sec:back:dS}

Let us assume that the background is a spatially flat de Sitter universe, with expansion rate $H$. The expansion is dominated by a positive cosmological constant in vacuum. 
The metric is written as follows, in Friedmann Lema\^itre Robertson Walker (FLRW) coordinates ($\tau,\rho,\theta,\phi$),
\begin{eqnarray}\label{d:desitter1}
&ds_{(0)}^{2} =-d\tau^2+e^{2H\tau}\big(d\rho^2+\rho^2d\Omega_2^2\big)\,,
\end{eqnarray}
and the linear background scalar field profile \cite{Babichev:2016jom,Sakstein:2016oel,Langlois:2017dyl} is, 
\begin{eqnarray}\label{d:desitter2}
\vp^{(0)}(\tau)&=&v_0\tau \,,
\end{eqnarray}
where, $v_0$ is a free constant coefficient and $d\Omega_2^2=d\theta^2+\sin^2\theta\, d\varphi^2$.

In reality, the background scalar profile can be non-linear if one accounts for the deviation from pure de Sitter space-time arising from the presence of matter. 
We leave the analysis of this point for future investigation. 

At the background level, the scalar field equation, Eq. \eqref{Ia*qDHOST}, reduces to 
\begin{equation}
 \partial_\tau(e^{3H\tau}J^\tau)=0\,.
\end{equation}
The solution, $J^\tau\sim e^{-3H\tau}$ approaches to zero very fast over time. Therefore, $J^\tau=0$ can be considered the acceptable particular solution. 
\begin{eqnarray}\label{eq:20}
2v_0\big(6 H^{2} \zeta_{1,X}^{(0)}v_0^{2} - 6 H^{2} \zeta_1^{(0)} - A\big) =0\,.
\end{eqnarray}

The subscript ${,X}$ denotes differentiation with respect to $X$ and the sub and superscript $(0)$ denotes the quantity in the de Sitter background profile. 
The background value of the kinetic term is 
\begin{equation}\label{X0}
X^{(0)}=-\,v_0^2\,,
\end{equation}
and we introduced the short-hand notation of the scalar function and its derivative on the background, as 
\begin{equation}\label{z0}
\zeta_1^{(0)}\equiv\zeta_1|_{X^{(0)}}\,,\qquad\zeta_{1,X}^{(0)}\equiv\zeta_{1,X}|_{X^{(0)}} \,.
\end{equation} 

The only non-vanishing and independent component\footnote{The other non-vanishing components of the background metric equation are $H_\rho\!^\rho=0$ and $H_\theta\!^\theta=0$ and $H_\phi\!^\phi=0$. However these component expressions are related together and to the scalar equation $J^\tau=0$ by the relation $H_\tau\!^\tau+v_0J^\tau=H_\rho\!^\rho=H_\theta\!^\theta=H_\phi\!^\phi$, so the only independent non-vanishing component of the background metric equation is $H_\tau\!^\tau=0$.} 
is that of the background metric equation $H_\tau\!^\tau=0$ , which implies 
 \begin{eqnarray}\label{eq:20.1} 
 12 H^{2} \zeta_{1,X}^{(0)}v_0^{4} - 18 H^{2} \zeta_1^{(0)}v_0^{2} - 6 H^{2} f + B \Lambda - A  v_0^{2}=0\,.
\end{eqnarray} 

One can find the relation of the free parameters from the background solution, Eq. \eqref{eq:20} and Eq. \eqref{eq:20.1},
\begin{equation}\label{AA}
A =6H^2\left(v_0^2\zeta_{1,X}^{(0)}-\zeta_{1}^{(0)}\right) \, ,
\end{equation}
and
\begin{equation}\label{constr}
\frac{f}{v_0^2}=\ddfrac{v_0^2\zeta_{1,X}^{(0)}-2\zeta_{1}^{(0)}}{1-B \sigma^2}\,.
\end{equation}
Let us introduce the dimensionless quantity $\sigma^2$, defined as
\begin{equation}\label{ffgf}
\sigma^2=\frac{\Lambda}{6H^2f},
\end{equation}
which depends on the model parameter $f$, which we assume to be positive. 

Vainshtein mechanism is studied in the spherical coordinate. Therefore, we now incorporate the previously obtained de Sitter background solution into the Schwarzschild-like coordinates ($t,r,\theta,\phi$), performing the transformation
\begin{equation}\label{eq:23}
\tau(t,r)= t +\frac{1}{2H}\ln\left(1-H^2r^2\right) \,,\qquad\qquad\rho(t,r)=\frac{r\,e^{-H t }}{\sqrt{1-H^2 r^2}}  \,,
\end{equation}
with $1-H^2r^2 > 0$ . 

Using the above transformations in Eq. \eqref{eq:23}, the cosmological background metric in \eqref{d:desitter1} and scalar profile \eqref{d:desitter2} can be rewritten in terms of the new Schwarzschild-like coordinates as 
\begin{eqnarray}\label{d:ds:sz:metric}
ds_{(0)}^{2} &=&  - \left(1-H^2 r^2\right) dt^2 + \frac{d\rho^2}{\left(1-H^2 r  ^2\right)}  + \rho^2 d\Omega_2^2 \,, \\
\vp^{(0)}(t,r)&=& v_0 t +\frac{v_0}{2H}\ln\left(1-H^2 r  ^2\right)\,. \label{d:ds:sz:scalar}
\end{eqnarray}

\section{Static spherically symmetric matter distribution in qDHOST class Ia*}
\label{sec:pert:qdhost}

Introducing static and spherically symmetric energy density, $\varepsilon$, and pressure, $P$, one gets the energy-momentum tensor 
\begin{equation}\label{eq:21}
T^\mu_\nu\equiv\mbox{diag}\Big\{-\varepsilon(r),P(r),P(r),P(r)\Big\}\,,
\end{equation}
which modifies the background space-time and the scalar profiles into
\begin{eqnarray}
&ds^2=-e^{\nu(r)}d t ^2+e^{\lambda(r)}d r  ^2+ r  ^2d\Omega_2^2 \,.
\label{eq:22}
\end{eqnarray}
Expressions \eqref{eq:22} include two radial-dependent metric potentials $\nu(r)$, $\lambda(r)$. We may neglect the time dependency in the perturbation of the scalar field in the sub-Horizon scale.
The scalar field equation, Eq. \eqref{Ia*qDHOST} leads to $J^r=0$ in the above coordinate system, which is expressed as
\begin{eqnarray}\label{gggf}
&&- 2 e^{- 3 \lambda  - \nu} \vp'\Big[r^{2} (v_{0}^{2} e^{\lambda } - e^{\nu} \vp'^{2})\Big]^{-1} \Big(A r^{2} v_{0}^{2} e^{3 \lambda  + \nu} - A r^{2} e^{2 \lambda  + 2 \nu} \vp'^{2} + 2 r v_{0}^{4} \zeta_{1X} e^{2 \lambda } \nu' \nonumber\\
&-& r v_{0}^{2} \zeta_{1} e^{2 \lambda  + \nu} \lambda'  - 3 r v_{0}^{2} \zeta_{1} e^{2 \lambda  + \nu} \nu' - 4 r v_{0}^{2} \zeta_{1X} e^{\lambda  + \nu} \nu' \vp'^{2} + 2 r \zeta_{1} e^{\lambda  + 2 \nu} \nu' \vp'^{2} \nonumber\\[1.0ex]
&+& 2 r \zeta_{1X} e^{2 \nu} \nu' \vp'^{4} - 2 v_{0}^{2} \zeta_{1X} e^{\lambda  + \nu} \vp'^{2} + 2 \zeta_{1} e^{\lambda  + 2 \nu} \vp'^{2} + 2 \zeta_{1X} e^{2 \nu} \vp'^{4}\Big)=0 \,.
\end{eqnarray}
In our convention the superscript $\,\!'$ denotes differentiation w.r.t. to the radial coordinate $r$.
The non vanishing metric equations are $H_t\!^t=0$ and $H_r\!^r=0$ lead to the following expressions for the energy density and pressure
\begin{eqnarray}\label{aaas1}
-\varepsilon&=&e^{- 3 \lambda  - \nu}\Big[r^{2} (- v_{0}^{2} e^{\lambda } + e^{\nu} \vp'^{2})\Big]^{-1} \Big(A r^{2} v_{0}^{4} e^{4 \lambda } - A r^{2} e^{2 \lambda  + 2 \nu} \vp'^{4} - B \Lambda r^{2} v_{0}^{2} e^{4 \lambda  + \nu} \nonumber\\
&+& B \Lambda r^{2} e^{3 \lambda  + 2 \nu} \vp'^{2} + 2 f  r v_{0}^{2} e^{3 \lambda  + \nu} \lambda'  - 2 f  r e^{2 \lambda  + 2 \nu} \lambda'  \vp'^{2} - 2 f  v_{0}^{2} e^{3 \lambda  + \nu} + 2 f  v_{0}^{2} e^{4 \lambda  + \nu} \nonumber\\[1.0ex]
&+& 2 f  e^{2 \lambda  + 2 \nu} \vp'^{2} - 2 f  e^{3 \lambda  + 2 \nu} \vp'^{2} + 4 r v_{0}^{4} \zeta_{1X} e^{2 \lambda } \lambda'  \vp'^{2} - 8 r v_{0}^{4} \zeta_{1X} e^{2 \lambda } \vp'\vp'' \nonumber\\[1.0ex]
&-& 10 r v_{0}^{2} \zeta_{1} e^{2 \lambda  + \nu} \lambda'  \vp'^{2} + 12 r v_{0}^{2} \zeta_{1} e^{2 \lambda  + \nu} \vp'\vp'' - 8 r v_{0}^{2} \zeta_{1X} e^{\lambda  + \nu} \lambda'  \vp'^{4} + 16 r v_{0}^{2} \zeta_{1X} e^{\lambda  + \nu} \vp'^{3}\vp'' \nonumber\\[1.0ex]
&+& 6 r \zeta_{1} e^{\lambda  + 2 \nu} \lambda'  \vp'^{4} - 8 r \zeta_{1} e^{\lambda  + 2 \nu} \vp'^{3}\vp'' + 4 r \zeta_{1X} e^{2 \nu} \lambda'  \vp'^{6} - 8 r \zeta_{1X} e^{2 \nu} \vp'^{5}\vp'' - 4 v_{0}^{4} \zeta_{1X} e^{2 \lambda } \vp'^{2} \nonumber\\[1.0ex]
&+& 6 v_{0}^{2} \zeta_{1} e^{2 \lambda  + \nu} \vp'^{2} + 4 v_{0}^{2} \zeta_{1X} e^{\lambda  + \nu} \vp'^{4} - 2 \zeta_{1} e^{\lambda  + 2 \nu} \vp'^{4}\Big) \,,
\end{eqnarray}
\begin{eqnarray}\label{aaas2}
P&=&e^{- 3 \lambda  - \nu}\Big[r^{2} (- v_{0}^{2} e^{\lambda } + e^{\nu} \vp'^{2})\Big]^{-1} \Big(- A r^{2} v_{0}^{4} e^{4 \lambda } + A r^{2} e^{2 \lambda  + 2 \nu} \vp'^{4} - B \Lambda r^{2} v_{0}^{2} e^{4 \lambda  + \nu} \nonumber\\
&+& B \Lambda r^{2} e^{3 \lambda  + 2 \nu} \vp'^{2} - 2 f  r v_{0}^{2} e^{3 \lambda  + \nu} \nu' + 2 f  r e^{2 \lambda  + 2 \nu} \nu' \vp'^{2} - 2 f  v_{0}^{2} e^{3 \lambda  + \nu} + 2 f  v_{0}^{2} e^{4 \lambda  + \nu}\nonumber\\[1.0ex]
&+& 2 f  e^{2 \lambda  + 2 \nu} \vp'^{2} - 2 f  e^{3 \lambda  + 2 \nu} \vp'^{2} - 4 r v_{0}^{4} \zeta_{1X} e^{2 \lambda } \nu' \vp'^{2} + 10 r v_{0}^{2} \zeta_{1} e^{2 \lambda  + \nu} \nu' \vp'^{2} \nonumber\\[1.0ex]
&+& 4 r v_{0}^{2} \zeta_{1} e^{2 \lambda  + \nu} \vp'\vp'' + 8 r v_{0}^{2} \zeta_{1X} e^{\lambda  + \nu} \nu' \vp'^{4} - 6 r \zeta_{1} e^{\lambda  + 2 \nu} \nu' \vp'^{4} - 4 r \zeta_{1X} e^{2 \nu} \nu' \vp'^{6} \nonumber\\[1.0ex]
&+& 2 v_{0}^{2} \zeta_{1} e^{2 \lambda  + \nu} \vp'^{2} + 4 v_{0}^{2} \zeta_{1X} e^{\lambda  + \nu} \vp'^{4} - 6 \zeta_{1} e^{\lambda  + 2 \nu} \vp'^{4} - 4 \zeta_{1X} e^{2 \nu} \vp'^{6}\Big) \,.
\end{eqnarray}

The matter source perturbs the metric potentials and the scalar field profiles about their cosmological values as
\begin{equation}\label{eq:24}
\nu( r  )\sim \nu^{(0)}( r  )+\delta\nu( r  )\,, 
\qquad\;\lambda( r  )\sim\lambda^{(0)}( r  )+\delta\lambda( r  )\,,
\;\quad\qquad\vp( r, t )\sim \vp^{(0)}( r  , t )+\delta\vp( r  ) \;,
\end{equation}
where $\delta\nu\ll\nu^{(0)}$, $\delta\lambda\ll\lambda^{(0)}$, $\delta\varphi\ll\varphi^{(0)}$. 
By definition, as distance approaches to the de Sitter horizon, these perturbations vanish and the background de Sitter solutions become important and approaches to the potentials and scalar field profile given in Eqs. \eqref{d:ds:sz:metric} and \eqref{d:ds:sz:scalar}. The kinetic energy and the scalar functions are perturbed as 
\begin{equation}\label{ddef}
X\sim X^{(0)}+\delta X\, ,\qquad\qquad\delta\zeta_1\sim\zeta_1^{(0)}+\delta\zeta_1 \;,
\end{equation}
where we have defined $\delta \zeta_1=\zeta_{1,X}^{(0)}\,\delta X$ and
\begin{equation}
\delta X=\left(\frac{1}{1-H^2r^2}\right)v_0^2\delta\nu- \left(\frac{H^2r^2}{1-H^2r^2}\right)v_0^2\delta\lambda-2 (Hr)v_0\delta\vp' \,.
\end{equation}
The background expression for $X^{(0)}$ is given in \eqref{X0}.

\subsection{Sub-Horizon non-relativistic Weak-Field Limit}
\label{subsec:subhorizon_weakfield}

The mass distribution of the matter source is
\begin{equation}
M(r)=4\pi\int_0^rx^2\varepsilon(x) \,dx \,.
\end{equation}

Obviously, $M(r\rightarrow \infty)=\mathcal{M}$, where $\mathcal{M}$ is the total mass of the structure, and $M(r=0)=0$.

There are three branches of solutions for $\delta \vp'$. The only solution where $\delta \vp$ decays at large scales, i.e., $\delta \vp' < 0 $, leads to the physical solution. In order to write the physical solution, one has to choose the weak-field limit, $\delta \nu \sim \delta \lambda \sim G_N M/r \ll 1$, and also consider the perturbation in the sub-horizon limit, $Hr\ll 1$. We keep all the non-linear terms of $\vp$ which dominate over the power of $\delta \nu $ and $\delta \lambda$. The $\delta \vp'^4 / v_0^4$ term is neglected with respect to 
$\delta \vp'^2 / v_0^2$ and $r \delta \vp' \delta \vp''/v^2$. 
We refer to Ref. \cite{Babichev:2016jom}, for a more detailed discussion of these approximations.

In the sub-horizon weak-field limit, $\delta X\sim v_0^2\delta \nu$ and the field equations, \eqref{gggf}, \eqref{aaas1}, and \eqref{aaas2}, become
\begin{eqnarray}
\label{Jr}
2r v_0^2 \zeta_{1}^{(0)}\delta\lambda'+2v_0^2r\left(3\zeta_1^{(0)}-2v_0^2\zeta_{1,X}^{(0)}\right)\delta\nu'-4\left(\zeta_1^{(0)}-v_0^2\zeta_{1,X}^{(0)}\right)\delta\vp'^2&=&0 \,,\\
\label{tt}
2f\left(\delta\lambda+r\delta\lambda'\right)+2\left(3\zeta_1^{(0)}-2v_0^2\zeta_{1,X}^{(0)}\right)\left(\delta\vp'^2+2r\delta\vp'\delta\vp''\right)-\frac{M'}{4\pi}&=&0 \,,\\
\label{rr}
2f\left(\delta\lambda-r\delta\nu'\right)+2\zeta_1^{(0)}\left(\delta\vp'^2+2r\delta\vp'\delta\vp''\right)+r^2P&=&0 \,.
\end{eqnarray}

Integrating \eqref{tt} one can obtain
\begin{equation}\label{gg}
2 f r \delta\lambda + 2 r\left(3  \zeta_1^{(0)}- 2  \zeta_{1,X}^{(0)}v_0^{2}\right) \delta\vp'^{2} - \frac{M}{4 \pi}+k=0 \,.
\end{equation}
The integrating constant, $k = 0$ under such physical system (because first three terms will be zero at the center), leaves
\begin{equation}\label{rtr}
\delta\lambda=- \frac{3 \zeta_1^{(0)}}{f} \delta\vp'^{2} + \frac{2 \zeta_{1,X}^{(0)}}{f} v_0^{2} \delta\vp'^{2} + \frac{M}{8\pi r f} \,.
\end{equation}
In the non-relativistic limit, $\varepsilon\gg P$, the combination of \eqref{rtr} and \eqref{rr} leaves
\begin{equation}\label{ppp}
\delta\nu'=\frac{2 \zeta_1^{(0)}}{f} \delta\vp'\delta\vp''- \frac{2 \zeta_1^{(0)}}{r f} \delta\vp'^{2} + \frac{2  \zeta_{1,X}^{(0)}v_0^{2}}{r f} \delta\vp'^{2} + \frac{M}{8\pi r^{2} f} \,.
\end{equation}
By substituting \eqref{rtr} and \eqref{ppp} into \eqref{Jr}, we may find three branches of solutions for $\delta \vp'$. The only physical branch, which decays at the large scale and leads to the asymptotic de Sitter expansion is 
\begin{equation}\label{dddfd}
\delta\vp'^2=- \frac{v_0^{2} \left( \zeta_1^{(0)}r M' + 2  \zeta_1^{(0)}M - 2  \zeta_{1,X}^{(0)}v_0^{2} M\right)}{16\pi r \left(-  \zeta_1^{(0)}+  \zeta_{1,X}^{(0)}v_0^{2}\right) \left(3  \zeta_1^{(0)}v_0^{2} - 2  \zeta_{1,X}^{(0)}v_0^{4} + f\right)}\,.
\end{equation}
In such a weak-field limit, the Schwarzschild potentials $\delta\lambda$ and $\delta\nu$ are related to the Newtonian potential and curvature perturbations by
\begin{equation}\label{aaa}
\frac{d\Phi(r)}{dr}=\frac{\delta\nu'(r)}{2}\,, \qquad\qquad \frac{d\Psi(r)}{dr}=\frac{\delta\lambda(r)}{2r} \,.
\end{equation}
Inserting the relation \eqref{dddfd} in \eqref{rtr} and \eqref{ppp} we find the potentials of the qDHOST model in terms of $G_N$, 
\begin{equation}\label{xxx1}
\frac{d\Phi(r)}{dr}=\frac{G_NM(r)}{r^2}+\frac{\Upsilon_1G_NM''(r)}{4}\,,
\end{equation}
\begin{equation}\label{xxx2}
\frac{d\Psi(r)}{dr}=\frac{G_NM(r)}{r^2}-\frac{5\Upsilon_2G_NM'(r)}{4r^2}\,,
\end{equation}
\noindent
where $G_N$, $\Upsilon_{1,2}$ parameters defined as 
\begin{equation}\label{aaa1}
G_N=\frac{1}{16\pi \left(3  \zeta_1^{(0)}v_0^{2} - 2  \zeta_{1,X}^{(0)}v_0^{4} + f\right)} \,,
\end{equation}
\begin{equation}\label{aaa2}
\Upsilon_1=- \frac{2 \zeta_1^{(0)\,2} v_0^{2}}{f\left(-  \zeta_1^{(0)}+  \zeta_{1,X}^{(0)}v_0^{2}\right)}\,,
\end{equation}
\begin{equation}\label{aaa3}
\Upsilon_2=\frac{2  \zeta_1^{(0)}v_0^{2} \left(- 3  \zeta_1^{(0)}+ 2  \zeta_{1,X}^{(0)}v_0^{2}\right)}{5 f\left(-  \zeta_1^{(0)}+  \zeta_{1,X}^{(0)}v_0^{2}\right)} \,.
\end{equation}

After using the background equations \eqref{constr} and restoring the proper dimensions to the parameters, we find
\begin{equation}\label{sss1}\displaystyle
G_N=\ddfrac{1}{2\bar{f}\left[\frac{3\zeta_1^{(0)}-2v_0^2\zeta_{1,X}^{(0)}}{2\zeta_1^{(0)}-v_0^2\zeta_{1,X}^{(0)}}(B\sigma^2-1)+1\right]}\,\mathcal{G} \,,
\end{equation}
\begin{equation}\label{sss2}
\Upsilon_1=\ddfrac{2\zeta_1^{(0)\,2}}{\left(\zeta_{1,X}^{(0)}v_0^2-\zeta_1^{(0)}\right)\left(\zeta_{1,X}^{(0)}v_0^2-2\zeta_1^{(0)}\right)}(B\sigma^2-1) \,,
\end{equation}
\begin{equation}\label{sss3}
\Upsilon_2=-\ddfrac{2\zeta_1^{(0)}\left(2\zeta_{1,X}^{(0)}v_0^2-3\zeta_1^{(0)}\right)}{5\left(\zeta_{1,X}^{(0)}v_0^2-\zeta_1^{(0)}\right)\left(\zeta_{1,X}^{(0)}v_0^2-2\zeta_1^{(0)}\right)}(B\sigma^2-1) \,.
\end{equation}
where we introduced the normalised $\bar{f}=fm_{pl}^{-2}$ where the Planck mass is defined as $m_{pl}^{-2}=8\pi\mathcal{G}$. $\mathcal{G}$ is the gravitational constant.

Non-vanishing $\Upsilon_{1,2}$ parameters in expressions \eqref{xxx1} and \eqref{xxx2} determine the breaking of Vainshtein screening. Inside the matter source, the radial dependency of mass suggests non-zero value of $M'(r)$ and $M''(r)$. However, the mass of the source is constant outside the source, hence $M'(r)=M''(r)=0$ and $\Upsilon_1=\Upsilon_2=0$, which confirm that one can recover GR outside extended sources within the Vainshtein radius, therefore, $\gamma_{PPN}=1$.

We have found that Vainshtein screening can be recovered inside the matter if $\Upsilon_1=\Upsilon_2=0$. The common factor in the numerator of the expressions of the $\Upsilon_1$ and $\Upsilon_1$ in Eq. \eqref{aaa2} and \eqref{aaa2} is $\zeta_1^{(0)} {\vp'_{(0)}}$. Therefore, $\Upsilon_{1,2}$ can set to zero in principle by setting either $\zeta_1^{(0)} = 0$ or ${\vp'_{(0)}}=0$. The second condition will kill the scalar degrees of freedom, thus making the model not interesting. It is possible to find such a non-zero function, $\zeta_1 (X)$, whose background value is zero, $\zeta_1^{(0)} = 0$, while the derivative of the function with respect to the kinetic energy in the background is non-zero, $\zeta_{1,X}^{(0)} \neq 0$. We have found that Vainshtein screening can be recovered fully within qDHOST, if we impose the condition, $\zeta_1^{(0)}=0$, on the free functions of the qDHOST theories, and the modified Newton constant becomes 
\begin{equation}
G_N=\frac{1}{2\bar{f}\big(2B\sigma^2-1\big)}\,\mathcal{G}\qquad\qquad\Upsilon_1=\Upsilon_2=0 \,.
\end{equation}

As a side note, we present a special case, where both, $\Upsilon_{1,2}$ are equal but non-zero, i.e., both the Newtonian potentials and the curvature perturbations break down the 
standard Newtonian 
behaviour in the same way, for the condition, $\zeta_1^{(0)}+v_0^2\zeta_{1,X}^{(0)}=0$ and we call it \textit{symmetric breaking of Vainshtein screening}. 
The expression of the $G_N$ is
\begin{equation}\label{fff}
G_N=\frac{3}{2\bar{f}\big(5B\sigma^2-2\big)}\,\mathcal{G}\,,\qquad\qquad\Upsilon_1=\Upsilon_2=-\frac{1}{3}\big(1-B\sigma^2\big)\,.
\end{equation}

For the specific choices of the parameters of qDHOST 
$\zeta_1\rightarrow f_4X\,, A\rightarrow-k_2\,, B\rightarrow1\,$, 
will lead to the restricted GLPV-beyond Horndeski class of theories studied in \cite{Babichev:2016jom}. 
One can obtain the same condition, $\zeta_1^{(0)}+v_0^2\zeta_{1,X}^{(0)}=0$\,, for the restricted GLPV-beyond Horndeski model and the full results and analysis can be recovered from the above expressions into the Eq. \eqref{fff} \cite{Babichev:2016jom}.

\section{Propagation of Gravitational Waves}
\label{sec:gw:dhost}
We are interested in the propagation of GW in a cosmological background. The tensor perturbation to the metric around the cosmological background is defined as 
\begin{equation}
g_{\mu\nu}=\left({\begin{array}{cc}
 -a^2 (\eta) & 0\\
 0 & a^2 (\eta)\delta_{ij} + 2 a^2 (\eta) h_{ij}
 \end{array}}
 \right)\,,
\end{equation}
where $h_{ij}$ is the tensor perturbation. We are using conformal time, $\eta$. The tensor perturbations are traceless and transverse, i.e., $h_{ii}=0=\partial^i h_{ij}$, where the latin indices $i,j$ refer to spatial coordinates.

The equation of motion for tensor modes in the DHOST Class Ia* reads
\begin{eqnarray}
 &&2f \nabla^2 h_{ij} + \ddot h_{ij} \left(-2f+ \frac{1}{a^4} \zeta_3 \dot \vp^4\right) + \dot h_{ij} \Bigg( -4 f\mathcal{H} + \frac{1}{a^4}\left(4\zeta _3 \dot \varphi^3 \ddot \varphi-2 \zeta _3 \mathcal{H} \dot\vp^4\right) \nn\\&&\quad+  \frac{1}{a^6} \left(2 \dot \zeta _3 \mathcal{H} \dot \varphi^6-2 \dot \zeta _3 \dot \varphi^5 \ddot \varphi\right)\Bigg)+h_{ij} \Bigg( -2 A \dot \varphi^2-2 B \Lambda  a^2+4 f \mathcal{H}^2+8 f\dot{\mathcal{H}}
   \nn\\&&\quad +\frac{1}{a^4}\left(14 \zeta _3 \mathcal{H}^2 \dot \varphi^4-4 \zeta _3 \dot {\mathcal{H}} \dot \varphi^4-16 \zeta _3 \mathcal{H} \ddot \varphi \dot \varphi^3\right) 
   +\frac{1}{a^6}\Big(8 \dot \zeta _3\mathcal{H} \dot \varphi^5 \ddot \varphi-8 \dot \zeta_3 \mathcal{H}^2 \dot \varphi^6 \Big)
   \Bigg)=0\,.\qquad
\end{eqnarray}
Dot denotes the derivative with respect to the conformal time, and $\mathcal{H} = \dot a / a$. As we seek solutions whose spatial dependence is given by $exp(i \vec{k} \cdot \vec{x})$, $\nabla^2 h_{ij}$ implies $-k^2h_{ij}$. The squared speed of GW propagation  is
 \begin{eqnarray}
 c_T^2 
 &=& \frac{f}{ f - \frac{1}{2a^2} \zeta_3^{(0)} {\dot \vp_{(0)}}^4}\,,\\
 &=& 1 - \frac{\zeta_1^{(0)} {\dot \vp_{(0)}}^2}{f + \zeta_1^{(0)} {\dot \vp_{(0)}}^2} \,, \label{e:ct:qdhost}\\
 &=&1+\alpha_T\,,
\end{eqnarray}
as $\zeta_3 = \frac{2}{X} \zeta_1$, $\zeta_3^{(0)} = - \frac{2a^2}{{\dot \vp_{(0)}}^2} \zeta_1^{(0)}$ and newly defined \cite{Bellini:2014fua}, $\alpha_T=- \frac{\zeta_1^{(0)} {\dot \vp_{(0)}}^2}{f + \zeta_1^{(0)} {\dot \vp_{(0)}}^2}$.

Therefore $c_T^2 = c^2 = 1$ when $\alpha_T$ vanishes. 

Note that the parameters in Eq. \eqref{e:ct:qdhost} depend on the background only \cite{deRham:2016wji}. It is interesting to note that the numerator of $\alpha_T$ is the same as in 
the expression of $\Upsilon_{1,2}$ given in Eq. \eqref{aaa2} and \eqref{aaa3}.
Therefore, similarly to the discussion on recovering screening in subsection \ref{subsec:subhorizon_weakfield}, $c_T^2 = c^2 = 1$ can be obtained in principle by setting $\zeta_1^{(0)} = 0$, without setting $\zeta_1 (X)=0$. Considering, however, the tight constraint arising from GW170817/GRB170817A, we need to ensure that even a small deviation 
of $\zeta_1^{(0)}$ in the background from zero will not lead to huge contributions to the $c_T^2$ or $\alpha_T$.

Eliminating $H^2$ from Eq. \eqref{eq:20} and \eqref{eq:20.1}, and then solving the equation for $v_0^2$, and use the expression of ${\vp'_{(0)}}^2$ given in \eqref{X0}
\begin{eqnarray}
{\vp'_{(0)}}^2 &=& - v_0^2\\
&=& \frac{ A f  + \Lambda B  \zeta_1^{(0)}}{2A \zeta_1^{(0)} - \Lambda B {\zeta_{1,X}}^{(0)}},
\label{e:ke:back:v}
\end{eqnarray}
and
\begin{eqnarray}
6 H^2 &=& \frac{ -2 A \zeta_1^{(0)}  + \Lambda B \zeta_{1,X}^{(0)} }{2{\zeta_1^{(0)}}^2 + f {\zeta_{1,X}}^{(0)}}.
\end{eqnarray}

Now, the extra quantity in the denominator of Eq. \eqref{e:ct:qdhost}, which contributes to the $c_T$ different than 1, will be same in the cosmic time (as it is dimensionless) and becomes
\begin{eqnarray}
&& \frac{\zeta_1^{(0)} {\vp'_{(0)}}^2}{f} \\ 
&=& \frac{ A + \Lambda B  \frac{\zeta_1^{(0)}}{f}}{2A  - \Lambda B \frac{{\zeta_{1,X}}^{(0)}}{\zeta_1^{(0)}}}\,. \label{e:alphat}
\end{eqnarray}

We can see from Eq. \eqref{e:alphat}, that a small deviation of $\zeta_1^{(0)}$ from zero will contribute by a large amount to $\alpha_T$, thus requiring a huge fine-tuning of the constants, $f$, $A$ and $B$. This conclusion would have been more complicated to reach, if $f$, $A$, and $B$ were functions instead of constants. Therefore, at least for the restricted shift-symmetric Ia* class qDHOST theory analysed here, one has to require $\zeta_1=0$ in order to satisfy $c_T^2 = 1$. 

In summary, the two expressions for the gravitational potentials in Eqs. \eqref{aaa2} and \eqref{aaa3} lead to exactly the same condition, in order to recover GR on small scales in Eq. \eqref{e:alphat}.
$\zeta_1=0$ implies that $\mathcal{L}_{\vp}=0$ in our action \eqref{ac:total}, corresponding to the Lagrangians that are quadratic in second-order derivatives of the scalar field. Thus these two independent constraints leave only GR plus $\mathcal{L}_{oth}$, which is k-essence field for the restricted qDHOST framework. On the other hand, some sector of the class Ia of qDHOST theories are surviving after the GW event; the Vainshtein screening for these classes is discussed in \cite{Crisostomi:2017lbg, Langlois:2017dyl}.

\section{Discussion and Conclusions}
\label{sec:concl}
In this article we studied the gravitational dynamics generated by a non-relativistic static and spherically symmetric cosmic structure within the framework of qDHOST theories of gravity. We restricted our study to the shift-symmetric Ia* class qDHOST gravitational model. We have explicitly deduced the covariant scalar-tensor field equations of qHOST and qDHSOT theories. We then studied a static, spherically symmetric cosmic structure embedded in de Sitter space-time for our qDHOST model. 
Similarly to the GLPV-beyond Horndeski theory, for the Ia* class of DHOST theory, the Vainshtein mechanism breaks down inside the cosmic structure, while GR can be recovered outside the matter source, within the Vainshtein radius. The expressions for the two gravitational potentials suggest us how to constrain the theory in order to recover GR within the Vainshtein radius. Then we explicitly derived the equation which governs the propagation of GWs in a cosmological background and found the conditions on the parameters of the theory, which allow to satisfy the $c_T^2 = c^2$ constraint. 
We then showed that the condition obtained from the screening of the fifth-force and the one on the GW propagation speed lead to the same constraint in Ia* class qDHOST theories.

\acknowledgments
We would like to thank F. Arroja, E. Bellini, P. Brax, A. Bombini, A. C. Davis, D. Langlois, S. Lanza, K. Koyama, and M. Zumalacarregui for useful discussions and comments on an early version of this manuscript. We thank to A. Dima and F. Vernizzi for sharing the draft of their paper before submission on the similar topic \cite{DimaVernizzi}. We would like to thank the PSI$^2$ DarkMod program where some useful feedbacks from participants while presenting a preliminary version of this work. PK acknowledges financial support from ``Fondazione Ing. Aldo Gini''. Some algebraic computations in this article were performed using Mathematica\footnote{https://www.wolfram.com/mathematica/}, the tensor computed algebra package xAct\footnote{http://www.xact.es/} and xPand\footnote{http://www.xact.es/xPand/} \cite{Pitrou:2013hga} and Cadabra software \cite{Peeters:2006kp}\cite{Peeters:2007wn}.

While completing our manuscript, two papers appeared on the arXiv \cite{Crisostomi:2017lbg, Langlois:2017dyl}, which consider related classes of models. As long as the analyses overlap we reach the same conclusions.


\begin{thebibliography}{10}

\bibitem{Perlmutter:1998np}
{\scshape Supernova Cosmology Project} collaboration, S.~Perlmutter et~al.,
  \emph{{Measurements of Omega and Lambda from 42 high redshift supernovae}},
  \href{https://doi.org/10.1086/307221}{\emph{Astrophys. J.} {\bfseries 517}
  (1999) 565--586}, [\href{https://arxiv.org/abs/astro-ph/9812133}{{\ttfamily
  astro-ph/9812133}}].

\bibitem{Riess:1998cb}
{\scshape Supernova Search Team} collaboration, A.~G. Riess et~al.,
  \emph{{Observational evidence from supernovae for an accelerating universe
  and a cosmological constant}},
  \href{https://doi.org/10.1086/300499}{\emph{Astron. J.} {\bfseries 116}
  (1998) 1009--1038}, [\href{https://arxiv.org/abs/astro-ph/9805201}{{\ttfamily
  astro-ph/9805201}}].

\bibitem{Hinshaw:2012aka}
{\scshape WMAP} collaboration, G.~Hinshaw et~al., \emph{{Nine-Year Wilkinson
  Microwave Anisotropy Probe (WMAP) Observations: Cosmological Parameter
  Results}}, \href{https://doi.org/10.1088/0067-0049/208/2/19}{\emph{Astrophys.
  J. Suppl.} {\bfseries 208} (2013) 19},
  [\href{https://arxiv.org/abs/1212.5226}{{\ttfamily 1212.5226}}].

\bibitem{Woodard:2006nt}
R.~P. Woodard, \emph{{Avoiding dark energy with 1/r modifications of gravity}},
  \href{https://doi.org/10.1007/978-3-540-71013-4_14}{\emph{Lect. Notes Phys.}
  {\bfseries 720} (2007) 403--433},
  [\href{https://arxiv.org/abs/astro-ph/0601672}{{\ttfamily
  astro-ph/0601672}}].

\bibitem{Burgess:2014lwa}
C.~P. Burgess and M.~Williams, \emph{{Who You Gonna Call? Runaway Ghosts,
  Higher Derivatives and Time-Dependence in EFTs}},
  \href{https://doi.org/10.1007/JHEP08(2014)074}{\emph{JHEP} {\bfseries 08}
  (2014) 074}, [\href{https://arxiv.org/abs/1404.2236}{{\ttfamily 1404.2236}}].

\bibitem{Woodard:2015zca}
R.~P. Woodard, \emph{{Ostrogradsky's theorem on Hamiltonian instability}},
  \href{https://doi.org/10.4249/scholarpedia.32243}{\emph{Scholarpedia}
  {\bfseries 10} (2015) 32243},
  [\href{https://arxiv.org/abs/1506.02210}{{\ttfamily 1506.02210}}].

\bibitem{Horndeski:1974wa}
G.~W. Horndeski, \emph{{Second-order scalar-tensor field equations in a
  four-dimensional space}},
  \href{https://doi.org/10.1007/BF01807638}{\emph{Int. J. Theor. Phys.}
  {\bfseries 10} (1974) 363--384}.

\bibitem{Gleyzes:2014dya}
J.~Gleyzes, D.~Langlois, F.~Piazza and F.~Vernizzi, \emph{{Healthy theories
  beyond Horndeski}},
  \href{https://doi.org/10.1103/PhysRevLett.114.211101}{\emph{Phys. Rev. Lett.}
  {\bfseries 114} (2015) 211101},
  [\href{https://arxiv.org/abs/1404.6495}{{\ttfamily 1404.6495}}].

\bibitem{Gleyzes:2014qga}
J.~Gleyzes, D.~Langlois, F.~Piazza and F.~Vernizzi, \emph{{Exploring
  gravitational theories beyond Horndeski}},
  \href{https://doi.org/10.1088/1475-7516/2015/02/018}{\emph{JCAP} {\bfseries
  1502} (2015) 018}, [\href{https://arxiv.org/abs/1408.1952}{{\ttfamily
  1408.1952}}].

\bibitem{Langlois:2015cwa}
D.~Langlois and K.~Noui, \emph{{Degenerate higher derivative theories beyond
  Horndeski: evading the Ostrogradski instability}},
  \href{https://doi.org/10.1088/1475-7516/2016/02/034}{\emph{JCAP} {\bfseries
  1602} (2016) 034}, [\href{https://arxiv.org/abs/1510.06930}{{\ttfamily
  1510.06930}}].

\bibitem{Crisostomi:2016czh}
M.~Crisostomi, K.~Koyama and G.~Tasinato, \emph{{Extended Scalar-Tensor
  Theories of Gravity}},
  \href{https://doi.org/10.1088/1475-7516/2016/04/044}{\emph{JCAP} {\bfseries
  1604} (2016) 044}, [\href{https://arxiv.org/abs/1602.03119}{{\ttfamily
  1602.03119}}].

\bibitem{Achour:2016rkg}
J.~Ben~Achour, D.~Langlois and K.~Noui, \emph{{Degenerate higher order
  scalar-tensor theories beyond Horndeski and disformal transformations}},
  \href{https://doi.org/10.1103/PhysRevD.93.124005}{\emph{Phys. Rev.}
  {\bfseries D93} (2016) 124005},
  [\href{https://arxiv.org/abs/1602.08398}{{\ttfamily 1602.08398}}].

\bibitem{BenAchour:2016fzp}
J.~Ben~Achour, M.~Crisostomi, K.~Koyama, D.~Langlois, K.~Noui and G.~Tasinato,
  \emph{{Degenerate higher order scalar-tensor theories beyond Horndeski up to
  cubic order}}, \href{https://doi.org/10.1007/JHEP12(2016)100}{\emph{JHEP}
  {\bfseries 12} (2016) 100},
  [\href{https://arxiv.org/abs/1608.08135}{{\ttfamily 1608.08135}}].

\bibitem{Zumalacarregui:2013pma}
M.~Zumalacárregui and J.~García-Bellido, \emph{{Transforming gravity: from
  derivative couplings to matter to second-order scalar-tensor theories beyond
  the Horndeski Lagrangian}},
  \href{https://doi.org/10.1103/PhysRevD.89.064046}{\emph{Phys. Rev.}
  {\bfseries D89} (2014) 064046},
  [\href{https://arxiv.org/abs/1308.4685}{{\ttfamily 1308.4685}}].

\bibitem{Langlois:2017mxy}
D.~Langlois, M.~Mancarella, K.~Noui and F.~Vernizzi, \emph{{Effective
  Description of Higher-Order Scalar-Tensor Theories}},
  \href{https://doi.org/10.1088/1475-7516/2017/05/033}{\emph{JCAP} {\bfseries
  1705} (2017) 033}, [\href{https://arxiv.org/abs/1703.03797}{{\ttfamily
  1703.03797}}].

\bibitem{Vainshtein:1972sx}
A.~I. Vainshtein, \emph{{To the problem of nonvanishing gravitation mass}},
  \href{https://doi.org/10.1016/0370-2693(72)90147-5}{\emph{Phys. Lett.}
  {\bfseries 39B} (1972) 393--394}.

\bibitem{Babichev:2013usa}
E.~Babichev and C.~Deffayet, \emph{{An introduction to the Vainshtein
  mechanism}},
  \href{https://doi.org/10.1088/0264-9381/30/18/184001}{\emph{Class. Quant.
  Grav.} {\bfseries 30} (2013) 184001},
  [\href{https://arxiv.org/abs/1304.7240}{{\ttfamily 1304.7240}}].

\bibitem{Brax:2004qh}
P.~Brax, C.~van~de Bruck, A.-C. Davis, J.~Khoury and A.~Weltman,
  \emph{{Detecting dark energy in orbit - The Cosmological chameleon}},
  \href{https://doi.org/10.1103/PhysRevD.70.123518}{\emph{Phys. Rev.}
  {\bfseries D70} (2004) 123518},
  [\href{https://arxiv.org/abs/astro-ph/0408415}{{\ttfamily
  astro-ph/0408415}}].

\bibitem{Kimura:2011dc}
R.~Kimura, T.~Kobayashi and K.~Yamamoto, \emph{{Vainshtein screening in a
  cosmological background in the most general second-order scalar-tensor
  theory}}, \href{https://doi.org/10.1103/PhysRevD.85.024023}{\emph{Phys. Rev.}
  {\bfseries D85} (2012) 024023},
  [\href{https://arxiv.org/abs/1111.6749}{{\ttfamily 1111.6749}}].

\bibitem{Narikawa:2013pjr}
T.~Narikawa, T.~Kobayashi, D.~Yamauchi and R.~Saito, \emph{{Testing general
  scalar-tensor gravity and massive gravity with cluster lensing}},
  \href{https://doi.org/10.1103/PhysRevD.87.124006}{\emph{Phys. Rev.}
  {\bfseries D87} (2013) 124006},
  [\href{https://arxiv.org/abs/1302.2311}{{\ttfamily 1302.2311}}].

\bibitem{Koyama:2013paa}
K.~Koyama, G.~Niz and G.~Tasinato, \emph{{Effective theory for the Vainshtein
  mechanism from the Horndeski action}},
  \href{https://doi.org/10.1103/PhysRevD.88.021502}{\emph{Phys. Rev.}
  {\bfseries D88} (2013) 021502},
  [\href{https://arxiv.org/abs/1305.0279}{{\ttfamily 1305.0279}}].

\bibitem{Kase:2013uja}
R.~Kase and S.~Tsujikawa, \emph{{Screening the fifth force in the Horndeski's
  most general scalar-tensor theories}},
  \href{https://doi.org/10.1088/1475-7516/2013/08/054}{\emph{JCAP} {\bfseries
  1308} (2013) 054}, [\href{https://arxiv.org/abs/1306.6401}{{\ttfamily
  1306.6401}}].

\bibitem{Kobayashi:2014ida}
T.~Kobayashi, Y.~Watanabe and D.~Yamauchi, \emph{{Breaking of Vainshtein
  screening in scalar-tensor theories beyond Horndeski}},
  \href{https://doi.org/10.1103/PhysRevD.91.064013}{\emph{Phys. Rev.}
  {\bfseries D91} (2015) 064013},
  [\href{https://arxiv.org/abs/1411.4130}{{\ttfamily 1411.4130}}].

\bibitem{Koyama:2015oma}
K.~Koyama and J.~Sakstein, \emph{{Astrophysical Probes of the Vainshtein
  Mechanism: Stars and Galaxies}},
  \href{https://doi.org/10.1103/PhysRevD.91.124066}{\emph{Phys. Rev.}
  {\bfseries D91} (2015) 124066},
  [\href{https://arxiv.org/abs/1502.06872}{{\ttfamily 1502.06872}}].

\bibitem{Saito:2015fza}
R.~Saito, D.~Yamauchi, S.~Mizuno, J.~Gleyzes and D.~Langlois, \emph{{Modified
  gravity inside astrophysical bodies}},
  \href{https://doi.org/10.1088/1475-7516/2015/06/008}{\emph{JCAP} {\bfseries
  1506} (2015) 008}, [\href{https://arxiv.org/abs/1503.01448}{{\ttfamily
  1503.01448}}].

\bibitem{Sakstein:2015zoa}
J.~Sakstein, \emph{{Hydrogen Burning in Low Mass Stars Constrains Scalar-Tensor
  Theories of Gravity}},
  \href{https://doi.org/10.1103/PhysRevLett.115.201101}{\emph{Phys. Rev. Lett.}
  {\bfseries 115} (2015) 201101},
  [\href{https://arxiv.org/abs/1510.05964}{{\ttfamily 1510.05964}}].

\bibitem{Sakstein:2015aac}
J.~Sakstein, \emph{{Testing Gravity Using Dwarf Stars}},
  \href{https://doi.org/10.1103/PhysRevD.92.124045}{\emph{Phys. Rev.}
  {\bfseries D92} (2015) 124045},
  [\href{https://arxiv.org/abs/1511.01685}{{\ttfamily 1511.01685}}].

\bibitem{Jain:2015edg}
R.~K. Jain, C.~Kouvaris and N.~G. Nielsen, \emph{{White Dwarf Critical Tests
  for Modified Gravity}},
  \href{https://doi.org/10.1103/PhysRevLett.116.151103}{\emph{Phys. Rev. Lett.}
  {\bfseries 116} (2016) 151103},
  [\href{https://arxiv.org/abs/1512.05946}{{\ttfamily 1512.05946}}].

\bibitem{Sakstein:2016ggl}
J.~Sakstein, H.~Wilcox, D.~Bacon, K.~Koyama and R.~C. Nichol, \emph{{Testing
  Gravity Using Galaxy Clusters: New Constraints on Beyond Horndeski
  Theories}}, \href{https://doi.org/10.1088/1475-7516/2016/07/019}{\emph{JCAP}
  {\bfseries 1607} (2016) 019},
  [\href{https://arxiv.org/abs/1603.06368}{{\ttfamily 1603.06368}}].

\bibitem{Babichev:2016jom}
E.~Babichev, K.~Koyama, D.~Langlois, R.~Saito and J.~Sakstein,
  \emph{{Relativistic Stars in Beyond Horndeski Theories}},
  \href{https://doi.org/10.1088/0264-9381/33/23/235014}{\emph{Class. Quant.
  Grav.} {\bfseries 33} (2016) 235014},
  [\href{https://arxiv.org/abs/1606.06627}{{\ttfamily 1606.06627}}].

\bibitem{Sakstein:2016oel}
J.~Sakstein, E.~Babichev, K.~Koyama, D.~Langlois and R.~Saito, \emph{{Towards
  Strong Field Tests of Beyond Horndeski Gravity Theories}},
  \href{https://doi.org/10.1103/PhysRevD.95.064013}{\emph{Phys. Rev.}
  {\bfseries D95} (2017) 064013},
  [\href{https://arxiv.org/abs/1612.04263}{{\ttfamily 1612.04263}}].

\bibitem{Sakstein:2017xjx}
J.~Sakstein and B.~Jain, \emph{{Implications of the Neutron Star Merger
  GW170817 for Cosmological Scalar-Tensor Theories}},
  \href{https://arxiv.org/abs/1710.05893}{{\ttfamily 1710.05893}}.

\bibitem{Crisostomi:2017lbg}
M.~Crisostomi and K.~Koyama, \emph{{Vainshtein mechanism after GW170817}},
  \href{https://arxiv.org/abs/1711.06661}{{\ttfamily 1711.06661}}.

\bibitem{Langlois:2017dyl}
D.~Langlois, R.~Saito, D.~Yamauchi and K.~Noui, \emph{{Scalar-tensor theories
  and modified gravity in the wake of GW170817}},
  \href{https://arxiv.org/abs/1711.07403}{{\ttfamily 1711.07403}}.

\bibitem{TheLIGOScientific:2017qsa}
{\scshape Virgo, LIGO Scientific} collaboration, B.~Abbott et~al.,
  \emph{{GW170817: Observation of Gravitational Waves from a Binary Neutron
  Star Inspiral}},
  \href{https://doi.org/10.1103/PhysRevLett.119.161101}{\emph{Phys. Rev. Lett.}
  {\bfseries 119} (2017) 161101},
  [\href{https://arxiv.org/abs/1710.05832}{{\ttfamily 1710.05832}}].

\bibitem{Monitor:2017mdv}
{\scshape Virgo, Fermi-GBM, INTEGRAL, LIGO Scientific} collaboration, B.~P.
  Abbott et~al., \emph{{Gravitational Waves and Gamma-rays from a Binary
  Neutron Star Merger: GW170817 and GRB 170817A}},
  \href{https://doi.org/10.3847/2041-8213/aa920c}{\emph{Astrophys. J.}
  {\bfseries 848} (2017) L13},
  [\href{https://arxiv.org/abs/1710.05834}{{\ttfamily 1710.05834}}].

\bibitem{deRham:2016wji}
C.~de~Rham and A.~Matas, \emph{{Ostrogradsky in Theories with Multiple
  Fields}}, \href{https://doi.org/10.1088/1475-7516/2016/06/041}{\emph{JCAP}
  {\bfseries 1606} (2016) 041},
  [\href{https://arxiv.org/abs/1604.08638}{{\ttfamily 1604.08638}}].

\bibitem{Lombriser:2015sxa}
L.~Lombriser and A.~Taylor, \emph{{Breaking a Dark Degeneracy with
  Gravitational Waves}},
  \href{https://doi.org/10.1088/1475-7516/2016/03/031}{\emph{JCAP} {\bfseries
  1603} (2016) 031}, [\href{https://arxiv.org/abs/1509.08458}{{\ttfamily
  1509.08458}}].

\bibitem{Bettoni:2016mij}
D.~Bettoni, J.~M. Ezquiaga, K.~Hinterbichler and M.~Zumalacárregui,
  \emph{{Speed of Gravitational Waves and the Fate of Scalar-Tensor Gravity}},
  \href{https://doi.org/10.1103/PhysRevD.95.084029}{\emph{Phys. Rev.}
  {\bfseries D95} (2017) 084029},
  [\href{https://arxiv.org/abs/1608.01982}{{\ttfamily 1608.01982}}].

\bibitem{Ezquiaga:2017ekz}
J.~M. Ezquiaga and M.~Zumalacárregui, \emph{{Dark Energy after GW170817: dead
  ends and the road ahead}},
  \href{https://arxiv.org/abs/1710.05901}{{\ttfamily 1710.05901}}.

\bibitem{Creminelli:2017sry}
P.~Creminelli and F.~Vernizzi, \emph{{Dark Energy after GW170817}},
  \href{https://arxiv.org/abs/1710.05877}{{\ttfamily 1710.05877}}.

\bibitem{Baker:2017hug}
T.~Baker, E.~Bellini, P.~G. Ferreira, M.~Lagos, J.~Noller and I.~Sawicki,
  \emph{{Strong constraints on cosmological gravity from GW170817 and GRB
  170817A}},  \href{https://arxiv.org/abs/1710.06394}{{\ttfamily 1710.06394}}.

\bibitem{Arai:2017hxj}
S.~Arai and A.~Nishizawa, \emph{{Generalized framework for testing gravity with
  gravitational-wave propagation. II. Constraints on Horndeski theory}},
  \href{https://arxiv.org/abs/1711.03776}{{\ttfamily 1711.03776}}.

\bibitem{Bellini:2014fua}
E.~Bellini and I.~Sawicki, \emph{{Maximal freedom at minimum cost: linear
  large-scale structure in general modifications of gravity}},
  \href{https://doi.org/10.1088/1475-7516/2014/07/050}{\emph{JCAP} {\bfseries
  1407} (2014) 050}, [\href{https://arxiv.org/abs/1404.3713}{{\ttfamily
  1404.3713}}].

\bibitem{DimaVernizzi}
A.~Dima and F.~Vernizzi, \emph{{Vainshtein Screening in Scalar-Tensor Theories
  before and after GW170817}}, {\emph{(to appear)} }.

\bibitem{Pitrou:2013hga}
C.~Pitrou, X.~Roy and O.~Umeh, \emph{{xPand: An algorithm for perturbing
  homogeneous cosmologies}},
  \href{https://doi.org/10.1088/0264-9381/30/16/165002}{\emph{Class. Quant.
  Grav.} {\bfseries 30} (2013) 165002},
  [\href{https://arxiv.org/abs/1302.6174}{{\ttfamily 1302.6174}}].

\bibitem{Peeters:2006kp}
K.~Peeters, \emph{{A Field-theory motivated approach to symbolic computer
  algebra}}, \href{https://doi.org/10.1016/j.cpc.2007.01.003}{\emph{Comput.
  Phys. Commun.} {\bfseries 176} (2007) 550--558},
  [\href{https://arxiv.org/abs/cs/0608005}{{\ttfamily cs/0608005}}].

\bibitem{Peeters:2007wn}
K.~Peeters, \emph{{Introducing Cadabra: A Symbolic computer algebra system for
  field theory problems}},
  \href{https://arxiv.org/abs/hep-th/0701238}{{\ttfamily hep-th/0701238}}.

\end{thebibliography}

\providecommand{\href}[2]{#2}\begingroup\raggedright\endgroup

\end{document}